\documentclass[twocolumn,prd,groupedaddress,nofootinbib]{revtex4}

\usepackage{graphicx}
\usepackage{dcolumn}
\usepackage{bm}
\usepackage{color}
\input{epsf}
\begin{document}

\title{AdS/CFT Correspondence and Strong Interactions}\thanks{Talk presented at the Fifth International Conference on Mathematical Methods in Physics, 24 - 28, April 2006, Rio de Janeiro, Brazil. To appear in the proceedings at http://pos.sissa.it/} 

\author{Henrique Boschi-Filho}
\email{boschi@if.ufrj.br}
\affiliation{Instituto de F\'{\i}sica, 
Universidade Federal do Rio de Janeiro, Caixa Postal 68528, 
RJ 21941-972 -- Brazil}
\author{Nelson R. F. Braga}
\email{braga@if.ufrj.br}
\affiliation{Instituto de F\'{\i}sica, 
Universidade Federal do Rio de Janeiro, Caixa Postal 68528, 
RJ 21941-972 -- Brazil}


\begin{abstract}  
The AdS/CFT correspondence is an exact duality between string 
theory in anti-de Sitter space and conformal field theories on its 
boundary. Inspired in this correspondence some relations between strings
and non conformal field theories have been found. 
Exact dualities in the non conformal case are intricate
but approximations can reproduce important physical results. 
A simple approximation consists in taking
just a slice of the AdS space with a size that can be related to the QCD energy scale.
Here we will discuss how this approach can be used to obtain the scaling
of high energy QCD scattering amplitudes, glueball masses and Regge trajectories, and the potential energy for a quark anti-quark pair. This set up is modified to include finite temperature effects and obtain a deconfining phase trasition.
\end{abstract}


\maketitle


\section{Introduction}

The AdS/CFT   correspondence is a duality between the large $N$ limit of $SU(N)$
superconformal field theories and string theory in a higher 
dimensional anti-de Sitter spacetime  \cite{Malda}.   
In this correspondence the AdS space shows up both as a near horizon 
geometry of a set of coincident D-branes. 
Precise prescriptions for the realization of the AdS/CFT   
correspondence were formulated in \cite{GKP,Wi}. 
In this correspondence, bulk fields act as classical sources for boundary 
correlation functions  \cite{Pe}$-$ \cite{FD}. 
One of the striking features  \cite{Wi} of this  correspondence is that it 
is a realization of the holographic principle  \cite{HOL1}$-$ \cite{BS}: 
``The degrees of freedom of  a quantum theory with gravity can be mapped 
on the corresponding boundary".
 
String theory was originally proposed as a model for strong 
interactions with the successful results of Veneziano amplitude in the Regge 
regime  \cite{V}$-$ \cite{SH}. 
However this model was unable to reproduce the experimental results for
hard and deep inelastic scatterings. 
These behaviours were latter explained by QCD which is today the accepted theory
 to describe strong interactions.
Despite of its success in high energies, QCD is non perturbative in 
the low energy regime, where one needs involved lattice calculations. 
In fact QCD and strings may be viewed as complementary theories.
This was first discovered by 't Hooft in the work  \cite{Planar} where he found a  relation between large $N$ $SU(N)$ gauge theories and strings  \cite{Pol}.

One of the long standing puzzles for the string description of strong interactions
was the high energy scattering at fixed angles.
For string theory in flat space such a process is soft: The amplitudes decay 
exponentially with energy  \cite{GSW} while both experimental 
data and QCD theoretical predictions  \cite{QCD1,BRO} indicate a hard behavior 
(amplitudes decaying with a power of energy.
A recent solution for this puzzle was  proposed by 
Polchinski and Strassler  \cite{PS}, inspired by the AdS/CFT   correspondence,
considering strings in a 10 dimensional space which assimptotically tends to an AdS$_5$ times a compact space. This space is considered as an approximation for the space 
dual to a confining gauge theory which can be associated with QCD.
An energy scale was introduced and identified with the lightest 
glueball mass. Then they found the glueball high energy scattering amplitude 
at fixed angles (with the correct QCD scaling) integrating the string amplitude 
over the warped AdS extra dimension weighted by the dilaton wave function.
Other approaches to this problem have also been discussed in  \cite{BB3}$-$ \cite{AN}. 

The first proposal to deform the AdS/CFT correspondence to describe strong interactions is due to Witten  \cite{Wi2}. In this work, inspired by Hawking and Page  \cite{Hawking:1982dh}, he suggests that the AdS space can acommodate a black hole (the AdS Schwarzschild space) and the horizon radius defines an energy scale which can be associated with the QCD mass gap. Following this proposal some authors have calculated Glueball masses in this space  \cite{MASSG}-\cite{MASSG7}. Other gauge/string dualities were discussed in  \cite{Maldacena:2000yy}-\cite{Polchinski:2000uf}.

Using the infrared cutoff proposed by Polschinski and Strassler we have calculated Glueball masses in an AdS slice from the zeros of Bessel functions  \cite{Boschi-Filho:2002vd,Boschi-Filho:2002ta}. This approach was later extended to higher spin states and applied to calculate  masses of light barions and mesons  \cite{deTeramond:2005su}. This description of higher spin states was then applied to obtain the Regge trajectories of Glueballs  \cite{Boschi-Filho:2005yh}, which showed to be consistent with the soft Pomeron trajactory.

Antoher successful application of the AdS/CFT correspondence is the calculation of Wilson loops. Using this duality one can identify the worldsheet area with the Wilson loop area of the correspponding gauge theory. Then, for static strings one finds the corresponding energy configuration. This idea has been implemented independently by Rey and Yee  \cite{RY}, and Maldacena  \cite{MaldaPRL}. It was later generalized to others spaces  \cite{Kinar:1998vq}. We have considered the calculation of Wilson loops in a AdS slice  \cite{Boschi-Filho:2005mw} finding the well known Cornell potential \cite{Quigg:1977dd}-\cite{Martin:1980jx} for a heavy quark anti-quark pair. This calculation was extended to finite temperature showing a deconfining phase trasition 
 \cite{Boschi-Filho:2006pe}. 

In this article we are going to review these recent results regarding approximate stringy descriptions of strong interactions. 
 
\section{Strings and Hard Scattering}

Let us start by briefly reviewing some early results on strong interactions
which lead to the proposal of string theory.
These results date back to the decade of 1960.
The first is the experimental observation, in hadronic scattering, of an apparently infinite tower of resonances with mass and angular momenta related by 
\begin{equation}
\label{1}
J \,\,\sim\,\,m^2 \,\alpha^\prime 
\end{equation}

\noindent where $\alpha^\prime \sim 1$ (GeV)$^{-2} $ is the Regge slope.  

Another important fact is that the properties of hadronic scattering 
in the so called Regge regime are nicely described by the amplitude 
postulated by Veneziano (in terms of Mandelstan variables ($s,t,u$)) 
\begin{equation}
\label{2}
 A(s,t) \,=\,{ \Gamma (-\alpha (s) )\, \Gamma (-\alpha (t) ) \over
 \Gamma (-\alpha (s) - \alpha (t) ) }
\end{equation}

\noindent where $ \Gamma (z)\,$ is the Euler gamma function and 
$ \alpha (s) \,=\,\alpha(0) + \alpha^\prime s $.

A strong motivation for relating hadrons to strings is the
fact that these two results of eqs. (\ref{1}) and (\ref{2}) can be reproduced 
from a relativistic bosonic string \cite{GSW,Polchinski:1998rq}.    
 After quantization, one finds that the spectrum of 
excitations shows up in representations satisfying eq. (\ref{1}). 
The string scattering amplitudes also reproduce the Veneziano result eq. (\ref{2}).

On the other hand, some problems concerning a possible string description
of hadronic physics have been a challenge for physicists for a long time.
One of them is the behavior of the scattering amplitudes at high energy.
If one considers the Regge limit, corresponding to $ s \rightarrow \infty \,,\,$ 
with fixed  $\, t \,:$ the Veneziano amplitude behaves as 
\begin{equation}
 A \sim s^{\alpha(t)}\,, 
\end{equation}

\noindent in agreement with experimental results. Actually this was one 
of the inputs for building up the Veneziano amplitude.
However considering high energy scattering at fixed angles that correspond to the
limit $ s \rightarrow \infty \,\,\, $ with $\,\,s/t\,$ fixed the amplitude behaves as
(soft scattering)
\begin{equation}
\,\,\, A_{_{Ven.}} \sim exp\,\{\, -\alpha^\prime s f (\theta)\, \} 
\end{equation}

\noindent while experimental results for hadrons show  a hard scattering behavior
\begin{equation}
\label{6}
 A_{exp.} \sim \, s^{(4-\Delta )/2} \,\,,
\end{equation}

\noindent that is reproduced from QCD \cite{QCD1,BRO}.

Recently Polchinski and Strassler \cite{PS} introduced an infrared cut off in 
the AdS space and reproduced the hard scattering behavior of strong interactions at fixed angles from string theory.
Inspired by the AdS/CFT correspondence they assumed 
a duality between gauge theory glueballs and string theory dilatons in an AdS slice
and found the QCD scaling.

This scaling was also obtained in  \cite{BB3} from a  mapping between quantum states 
in AdS space and its boundary found in  \cite{BB2}. 
We considered an AdS slice as approximately dual to a confining gauge 
theory.  The slice corresponds to the metric
\begin{equation}
\label{AdSPoincare}
ds^2=\frac {R^2 }{( z )^2}\Big( dz^2 \,+(d\vec x)^2\,
- dt^2 \,\Big)\,,
\end{equation}

\noindent with $\, 0\le z \le  z_{max} \,\, \sim 1/\mu $  where  $\mu$
is an energy scale chosen as the mass of the lightest glueball.
We used a mapping between Fock spaces of a scalar field in AdS space and operators 
on the four dimensional boundary, defined in  \cite{BB2}. 
Considering a scattering of two particles into $m$ particles 
one finds a relation between bulk and boundary scattering amplitudes \cite{BB3}
\begin{equation}
S_{Bulk} \, \sim \,  
 S_{Bound.} \,\,\Big( {\sqrt{\alpha^\prime} \over \mu }\Big)^{m+2} \,\, K^{(m+2)(1+d)}
\end{equation}

\noindent where $d $ is the scaling dimension of the boundary operators and $K$ is the boundary momentum scale. This leads to the scaling of the amplitude  
\begin{equation}
A_{Boundary} \,\sim \,s^{(4  - \Delta)/2 } \,,
\end{equation}

\noindent where $\Delta=(m+2)d$ is the total dimension of scattered particles. 
This reproduces the QCD scaling of eq. (\ref{6}). 
For some other results concerning QCD scattering properties from string theory see also 
 \cite{PS2}-\cite{AN3}. 
 
\section{Scalar glueball masses }

Using the phenomenological approach of introducing an energy scale 
by considering an AdS slice we found estimates for scalar glueball mass ratios 
 \cite{Boschi-Filho:2002vd,Boschi-Filho:2002ta}. 
In the AdS$_5$ slice we considered dilaton fields satisfying Dirichlet boundary conditions at $ z = z_{max}$ 
\begin{eqnarray}
\label{QF}
\Phi(z,\vec x,t) &=& \sum_{p=1}^\infty \,
\int { d^3 k \over (2\pi)^{3}}\,
{z^{2} \,J_2 (u_p z ) \over z_{max}\,\, w_p(\vec k ) 
\,J_{3} (u_p z_{max} ) }\nonumber\\
&\times& \lbrace { {\bf a}_p(\vec k )\ }
 e^{-iw_p(\vec k ) t +i\vec k \cdot \vec x}\,
\,+\,\,h.c.\rbrace\,,\nonumber
\end{eqnarray}

\noindent where $\,w_p(\vec k ) \,=\,
\sqrt{ u_p^2\,+\,{\vec k}^2}\,$, 
\begin{equation}
 u_p \,=\,\frac{ \chi_{_{2\,,\,p}}}{z_{max}} \,,
\end{equation}

\noindent is the momentum associated with the $z$ direction
and $\chi_{_{2\,,\,p}} $ are the zeroes of the Bessel functions:  
$\,J_2 (\chi_{_{2\,,\,p}} )=0\,.$
 
 On the boundary ($ z = 0)$ we considered scalar glueball states  $J^{PC}\,=\,0^{++}$ 
and their excitations $0^{++\ast},\,\,0^{++\ast\ast}$  with masses 
$\mu_p\,,\,p=1,2,...$. 
Assuming an approximate gauge/string duality the 
glueball masses are taken as proportional to the dilaton discrete modes: 
$$
{u_p \over \mu_p }\,=\,{\rm const.}\,
$$

\noindent So, the ratios of glueball masses are related  
to zeros of the Bessel functions
$$
{ \mu_p\over \mu_1 }\,=\,{\chi_{2\,,\,p}\over \chi_{2\,,\,1}}\,\,.
$$

\bigskip

\noindent Note that these ratios are independent of the size of the slice   $\,\,\,\,\,z_{max}\,$. 
Our estimates compared with $SU(3) $ lattice  \cite{LAT1,LAT2} and AdS-Schwarzschild 
 \cite{MASSG} (in GeV) are shown in Table I.

\begin{table}[htbp]
\centering
\begin{tabular}{|l|c|c|c|}
\hline
4d State  & {  lattice}, $N=3$ &
{  AdS-BH} & {  AdS slice} \\
 \hline
 $0^{++}$ & $1.61 \pm 0.15$   & 1.61  & 1.61  \\
 $0^{++*}$ &  2.8   & 2.38 & 2.64 \\
 $0^{++**}$ &   - & 3.11 & 3.64 \\
 $0^{++***}$ &  -  & 3.82 & 4.64\\
 $0^{++****}$ &  -  & 4.52 & 5.63\\
 $0^{++*****}$ &  -  & 5.21 & 6.62\\
\hline
\end{tabular}
\caption{Four dimensional glueball masses in GeV with Dirichlet boundary conditions. The value 1.61 of the third and fourth columns is an input taken from lattice results.}
\end{table}

\begin{table}[htbp]
\centering
\begin{tabular}{|l|c|c|c|c|}
\hline
3d State & {lattice, $N=3$} & 
{lattice,} {$N\rightarrow\infty$} &
{AdS-BH} & {AdS slice} \\
 \hline
 $0^{++}$ & $4.329 \pm 0.041$ & $4.065 \pm 0.055$ & 4.07 
& 4.07  \\
 $0^{++*}$ & $6.52 \pm 0.09$ & $6.18 \pm 0.13$ & 7.02 & 7.00\\
 $0^{++**}$ & $8.23 \pm 0.17$ & $7.99 \pm 0.22$ & 9.92 & 9.88 \\
 $0^{++***}$ &  - & - & 12.80 & 12.74 \\
 $0^{++****}$ &  - & - & 15.67 & 15.60\\
 $0^{++*****}$ & -  & - & 18.54 & 18.45\\
\hline
\end{tabular}
\caption{Three dimensional glueball masses in units of string tension with Dirichlet boundary conditions. The value 4.07
is an input from lattice.}
\end{table}

A similar approach was used also for glueball masses in QCD$_3$, taken as dual to 
scalar fields is AdS$_{4}$. In this case the Bessel functions are $J_{3/2} $ 
and  the mass ratios take the form 
\begin{equation}
{ \mu_p\over \mu_1 }\,=\,{\chi_{3/2\,,\,p}\over \chi_{3/2\,,\,1}}\,\,.
\end{equation}
 
\noindent Our results for QCD$_3$ are shown in Table II, again 
compared with lattice  \cite{LAT1,LAT2} and AdS-Schwarzschild 
 \cite{MASSG} results.
 For some other results concerning glueball masses using gauge/string dualities 
see for instance  \cite{Caceres:2000qe}-\cite{Caceres:2005yx}.

\section{Higher spin states and Regge trajectories}

Recently, very interesting results for the hadronic
spectrum were obtained by Teramond and Brodsky \cite{deTeramond:2005su} considering scalar, vector and fermionic fields in a sliced $ AdS_5 \times S^5 $ space.
It was proposed that massive  bulk states corresponding to fluctuations about 
the $AdS_5$ metric are dual to QCD states with  
angular momenta (spin) on the four dimensional boundary.
This way the spectrum of light baryons and mesons has been reproduced
from a holographic dual to QCD inspired in the AdS/CFT 
correspondence.

We used a similar approach to estimate masses of glueball states 
with different spins \cite{Boschi-Filho:2005yh}. 
The motivation was to compare the glueball Regge trajectories with
the pomeron trajectories. For soft pomerons  \cite{Landshoff:2001pp} experimental
results show that
\begin{equation}
\label{11}
J \,\approx \,  1.08 \,+\, 0.25\, M^2\,\,,
\end{equation}

\noindent where $M$ is the mass of the state in GeV. It is conjectured that the soft pomerons may be related to glueballs.
Recent lattice results are consistent with this interpretation \cite{Meyer:2004jc}.
 
We assumed that massive scalars in the AdS slice with mass $\mu$ 
are dual to boundary gauge theory states with spin $J$ related by: 
\begin{equation}
( \mu R )^2 \,=\, J ( J + 4 ) \,\,.
\end{equation}

We considered both Dirichlet and Neumann boundary conditions and the results for the
four dimensional glueball masses with even spin are shown in tables III and IV respectively.

\begin{table}
\centering
\begin{tabular}{ | c | c | c | c |} 
\hline 
Dirichlet $\,\,$ 
 & $\,\,$ lightest $\,\,$   &
$1^{st}$ excited $\,\,$ & $2^{nd}$ excited \\
glueballs $\,\,$     &   state & state & state \\
 \hline
 $0^{++}$ & 1.63  &  2.67 & 3.69 \\ 
 $2^{++}$ & 2.41    & 3.51  & 4.56  \\
 $4^{++}$ & 3.15  & 4.31  & 5.40  \\
 $6^{++}$ & 3.88  & 5.85  & 6.21 \\
 $8^{++}$ & 4.59 & 5.85  & 7.00 \\
 $10^{++}$ & 5.30 & 6.60   & 7.77 \\
\hline
\end{tabular}
\caption{     Higher spin glueball masses in GeV with Dirichlet boundary condition. 
The value 1.63 is an input from lattice.}
\end{table}

\begin{table}
\centering
\begin{tabular}{ | c | c | c | c |} 
\hline 
Neumann $\,\,$ 
 & $\,\,$ lightest $\,\,$   &
$1^{st}$ excited $\,\,$ & $2^{nd}$ excited \\
glueballs $\,\,$     &   state & state & state \\
 \hline
 $0^{++}$ & 1.63  & 2.98  & 4.33  \\ 
 $2^{++}$ & 2.54    & 4.06 & 5.47  \\
 $4^{++}$ & 3.45 & 5.09  & 6.56 \\
 $6^{++}$ & 4.34  & 6.09 & 7.62 \\
 $8^{++}$ & 5.23 & 7.08 & 8.66 \\
 $10^{++}$ & 6.12 & 8.05  & 9.68 \\
\hline
\end{tabular}
\caption{  Higher spin glueball masses in GeV with Neumann boundary condition. 
The value 1.63 is an input from lattice.  }
\end{table} 
   
We found non linear relations between spin and mass squared.
Considering a linear fit representing Regge trajectories
\begin{equation}
 J \,=\, \alpha_0 \,+\,\alpha^{\prime} \, M^2\,,
\end{equation}

\noindent we found for Neumann boundary conditions and the states  
$J^{++}\,$ with $J = 2,4,...,10\,$ 
\begin{equation}
\alpha^{\prime}\,=\,(\,0.26 \pm 0.02 \,)GeV^{-2} 
\qquad ; \qquad \alpha_0 \,=\,0.80 \pm 0.40\,,
\end{equation}

\noindent as shown in Figure 1.

\begin{figure}[htbp]
\begin{center}
\includegraphics[width=8cm]{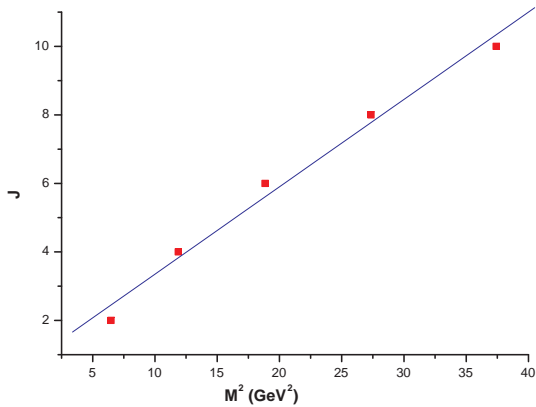}
\caption{  Spin versus mass squared for the lightest glueball states with 
Neumann boundary conditions from table IV. The line corresponds to the linear fit.}
\end{center}
\end{figure}

\begin{figure}[htbp]
\begin{center}
\includegraphics[width=8cm]{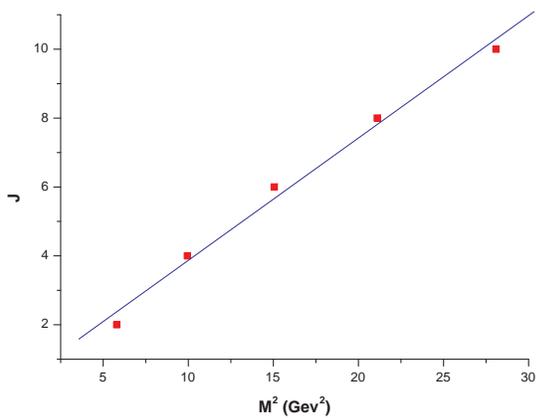}
\caption{ Spin versus mass squared for the lightest glueball states with 
 Dirichlet boundary conditions from table III. 
The line corresponds to the linear fit. }
\end{center}
\end{figure}
\noindent For Dirichlet boundary conditions, taking the states 
$J^{++}\,$ with $J = 2,4,...,10\,$ 
we found 
\begin{equation}
\alpha^{\prime}\,=\,( \, 0.36 \pm  \,0.02\,)\,GeV^{-2}\, \qquad ; \qquad
 \alpha_0 \,=\,0.32 \pm 0.36 \,,
\end{equation}

\noindent as shown in Figure 2.

So, Neumann boundary conditions give a glueball trajectory consistent with
that of pomerons, eq. (\ref{11}). These kind of boundary conditions appear in the Randall Sundrum 
model \cite{Randall:1999ee} as a consequence of the orbifold condition.



\section{Wilson loops and quark anti-quark potential}

Wilson loops are an important tool to discuss confinement in gauge theories since they 
show the behaviour of the energy associated with a given field configuration.
In the AdS/CFT correspondence Wilson loops for a heavy quark anti-quark pair in the
gauge theory can be calculated from a static string in the 
AdS space \cite{RY,MaldaPRL}. The corresponding energy is a non confining Coulomb
potential as expected for a conformal theory. For an excellent review and extension to other metrics see  \cite{Kinar:1998vq}.  
We calculated Wilson loops for a quark anti-quark pair in D3-brane space finding different behaviors, with respect to confinement, depending on the quark position \cite{Boschi-Filho:2004ci}. 

Here, we are going to review \cite{Boschi-Filho:2005mw} that the Wilson loop calculated in an AdS slice leads to the heavy quark anti-quark Cornell potential:
\begin{equation}
\label{Cornell}
E_{Cornell}(L) \,=\, -\frac{4}{3} \frac{a}{L} \,+\, \sigma L\,+\, constant\,\,,
\end{equation}
\noindent where $a = 0.39\,$ and $\sigma = 0.182$ GeV$^2\,$ \cite{Quigg:1977dd}-\cite{Martin:1980jx}. 

The metric (\ref{AdSPoincare}) of the AdS space can be rewritten as 
\begin{equation}
\label{metric}
ds^2 \,=\, \Big( {r^2\over R^2} \Big) ( -dt^2 + d{\vec x}^2 ) +  
\Big( {R^2\over r^2} \Big) dr^2 \,,
\end{equation}

\noindent where $\, r = R^2/z $. We have calculated the energy of a static string in an AdS slice defined by  $ r_2 \le r \le r_1 \,$. The quark anti-quark pair (string endpoints) is located at $r = r_1 $, separated by a four dimensional ($x$ coordinates) distance $L$  and there is an infrared cut off in the space at $ r= r_2$. From now on we choose $r_2 \,=\, R$. Note that there are two kinds of geodesics, as shown in figure 3, depending on the value of $L$. For small quark separation $ L \le L_{crit}\,$ the geodesics are curves (like curve {\bf a} ) with one minimum value of the coordinate $r = r_0 $ which is related to $L$ by 
\begin{equation}
\label{Lr1}
L (r_0 ) \,=\,\frac{ 2 R^2 }{r_0}\,I_1 (r_1/r_0 )
\end{equation}

\noindent where $I_1 (\xi )$ is the elliptic integral
\begin{equation}
I_1 (\xi )\,=\,\int_1^{\,\xi}\,
\frac{ d \rho }{\rho^2 \,\sqrt{ \rho^4 -1}}\,.
\end{equation}

\noindent The critical value corresponds to $ L_{crit} = L ( r_0 = R ) $ as in curve {\bf b} of figure 3.  

The energy for $L \le L_{crit}\,$ can be calculated as
\begin{equation}
\label{E-}
E^{\,(-)} \,=\, \frac{ 2 R^2 }{\pi \alpha^\prime} \,\frac{I_1 (r_1/r_0 )}{L}
\Big[ \, I_2 (r_1/r_0)\,-\,1 \Big] \,.
\end{equation}

\noindent where $1/2\pi \alpha^\prime $ is the string tension and we have subtracted the 
constant $ r_1 /\pi \alpha^\prime\,$  in such a way that the energy is finite 
even in the limit $r_1 \to \infty $. The integral $I_2 $ is  
\begin{equation}
I_2 (\xi) \,=\, \int_1^{\xi}\,
\Big[\,\frac{ \rho^2 }{\sqrt{ \rho^4 -1}} \,-\,1 \,\Big] d\rho \,.
\end{equation}

For $ L > L_{crit} $, the geodesics reach the infrared brane as shown in curve {\bf c} of figure 3.

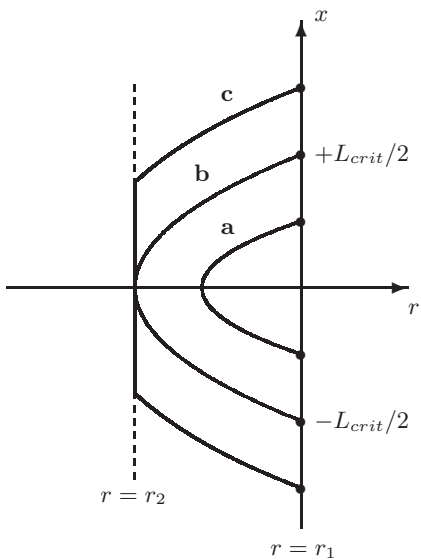
\begin{figure}
\centering
\setlength{\unitlength}{0.07in}
\vskip 5.5cm
{\begin{picture}(0,0)(18,0)
\rm\thicklines\bf
\put(25,-8){\vector(0,0){38}}
\put(26,-0.5){$- L_{crit}/2$}
\put(26,19.5){$+L_{crit}/2$}
\put(26,30){$x$}
\put(24.5,-0.5){$\bullet$}
\put(24.5,19.5){$\bullet$}
\put(24.5,14.5){$\bullet$}
\put(24.5,4.5){$\bullet$}
\put(24.5,24.5){$\bullet$}
\put(24.5,-5.5){$\bullet$}
\put(17,18){b}
\put(19,14){a}
\put(19,24){c}
\put(3,10){\vector(1,0){30}}
\put(33,8){$r $}
\put(22.7,-10 ){$r = r_1 $}
\put(10,-6){$r = r_2$}
\bezier{600}(25,15)(10,10)(25,5)
\bezier{600}(25,20)(0,10)(25,0)
\bezier{600}(25,25)(17,22)(12.6,18)
\bezier{600}(25,-5)(17,-2)(12.6,2)
\bezier{600}(12.5,18)(12.5,10)(12.5,2)
\bezier{600}(12.55,18)(12.55,10)(12.55,2)
\multiput(12.55,-4.3)(0,1){30}{\line(0,1){0.5}}
\end{picture}}
\vskip 2.5cm
{\caption{ Schematic representation 
of geodesics in the AdS slice. }}
\end{figure}

The energy can be calculated again subtracting the constant $ r_1 /\pi \alpha^\prime\,,$ associated with the quark masses. 
We obtain
\begin{eqnarray}
E^{\,(+)} = \frac{ R }{\pi \alpha^\prime} 
\Big[ I_2 (r_1/R) - 1 \,\Big]
\,+\,\frac{ 1 }{2 \pi \alpha^\prime\,}\,( L - L_{crit})\,  \cr\cr
= \frac{ R }{\pi \alpha^\prime} 
\Big[ I_2 (r_1/R)  - I_1 (r_1/R) - 1\,\Big]
+\frac{ L }{2 \pi \alpha^\prime} ,
\label{Er1}
\end{eqnarray}

\noindent where we have used the definition of $L$, eq. (\ref{Lr1}), with the identification that at $L=L_{crit}$ one has $r_0=r_2=R$. 
Taking the limit $ r_1 \to \infty \,$, we 
find a potential which is approximately Coulombian for small $ L $ and 
has a leading linear confining behaviour for  large distances
\begin{equation}
 E^{\,(+)} \,\sim \, \frac{1}{2\pi\alpha'} \, L \,.
\end{equation}

\noindent The identification of the static string energy with the Cornell potential (\ref{Cornell}) leads to 
\begin{equation}
a \,=\, \frac{3 C_1^2 R^2}{2\pi \alpha'}\,\,\,\,\,;\,\,\,\,\,\,\,
\sigma \,=\, \frac{1}{2\pi\alpha'}  
\end{equation} 
  
\noindent with $C_1\,= \,\sqrt{2}\pi^{3/2}/[\Gamma(1/4)]^2.$ So, for the particular choice $r_2=R$,  we find an effective AdS radius $ R~=~1.4~\,{\rm GeV}^{-1}\,$. 
Then the energies take the forms:\break
\begin{eqnarray} 
E^{\,(-)} &=&
\,-\, \frac{ 4 a }{ 3 \,L } \,
\,;  \qquad \hskip 1.3cm L \le L_{crit} \\ 
 E^{\,(+)} &=& -\, 4\sqrt{\frac{\,  \,a \,\sigma }{ 3\, }} \,+\,
 \sigma  L \,\,\,; \qquad\hskip .35cm L \ge L_{crit}
\end{eqnarray}
The shape of this potential energy is very close to the Cornell potential, as shown in figure~4.

\begin{figure}
\centering
\includegraphics[width=8cm]{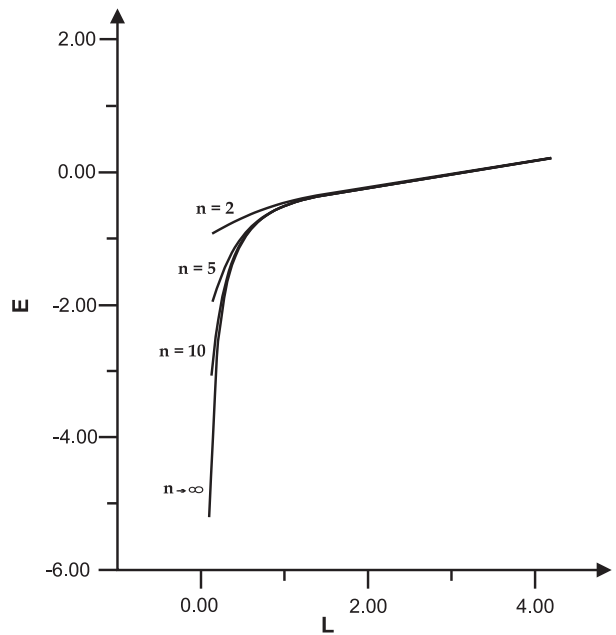}
{\caption{ Energy in GeV as a function of string end-points 
separation $L$ in GeV$^{-1}$, for AdS slices with
 $r_1 = nR $ and $r_2 = R $ . For $ n \to \infty$  
the energy behaves as the Cornell potential.}}
\end{figure}



\section{Quark anti-quark potential at finite temperature}  

A gauge/string duality involving a gauge theory at finite temperature 
was proposed by Witten in  \cite{Wi2} inspired by the work of
Hawking and Page \cite{Hawking:1982dh}.
In this approach, for high temperatures, the AdS space accommodates a  Schwarzschild black hole and the
horizon radius is proportional to the temperature. For low temperatures the dual space would be an AdS space with compactified time dimension, known as thermal AdS. 
The gauge string duality using the AdS Schwarzschild space has recently been applied to obtain the viscosity of a quark gluon plasma \cite{Policastro:2001yc,Kovtun:2004de}. 

An AdS Schwarzschild black hole metric was considered in \cite{Boschi-Filho:2006pe} as a phenomenological model for a space dual to a theory with both mass scale and finite temperature. The corresponding metric is
\begin{equation}
\label{metric}
ds^2 \,=\, \Big( {r^2\over R^2} \Big) ( -\, f(r) \,dt^2 \,+ \, d{\vec x}^2 \,) +  
\Big( {R^2\over r^2} \Big) \,\frac{dr^2}{f(r) } \,  \,+\, R^2 d^2 \Omega_5\,,
\end{equation}

\noindent where $r_2 \le r < \infty \,\,$, $\,\,\,  f(r)\,=\, 1\,-\, r_T^4 / r^4 \,\ $ 
and the Hawking temperature $T$ is related to the horizon radius by  $r_T\,=\, \pi\, R^2 \,T\,$. At zero temperature this space becomes an  Anti-de Sitter (AdS) slice. 
The problem of static strings in a space with metric (\ref{metric}), without any  cut off, was discussed in detail in \cite{Rey:1998bq,Brandhuber:1998bs}. 

Calculating the energy of static strings in this space we found a deconfinement phase transition at a critical temperature $T_C$. This transition shows up because, depending on the horizon radius (temperature) relative to the cut off position $r_2$, we have different behaviors for the energy. If $r_T \ge r_2$ the  string will not be affected by the presence of the brane since it does not cross the horizon. So the energy will be that described in ref.  \cite{Rey:1998bq,Brandhuber:1998bs} and there will be no confinement. If $r_T <  r_2$ the energy for large quark anti-quark distance $L$ will grow up linearly with $L$ with a temperature dependent coefficient and the quarks are confined. 
The critical temperature $\, T_C\,$ corresponds to $r_T \,=\, r_2$. 

As in the zero temperature case we choose $r_2 = R$ from now on. 
We show in figure 5 the energies obtained in  \cite{Boschi-Filho:2006pe}
for temperatures $ T =0,\, T = 0.8 T_C \,,\, T = T_C\,$ and $ \,T = 2 T_C\,$. 
This figure illustrates the fact that in our model the energy of static strings associated with the quark anti-quark potential present a confining behavior for 
temperatures below  $ T_C \,=\, 1 /\, \pi\, R\,$. In this case there is a linear term in the energy, for large quark distances $L\,\,$ given by  $\,E \sim  \sigma (T)\,L\,$ with 
\begin{equation}
\label{tensionT}
\sigma (T) \,=\,\frac{1}{2 \pi \alpha^\prime} \sqrt{ 1 -  (\pi R T )^4 }\,\,\hskip1cm ( T < T_C ).
\end{equation}
 
\noindent At zero temperature this coefficient is identified with the string tension of the Cornell potential 
$ 1/ ( 2 \pi \alpha^\prime ) \,=\, 0.182$ Gev$^2$. 
For temperatures $ T \ge  T_C $ there is no confinement since the energy is finite 
when $L \to \infty$.   
Choosing the brane position to have the value $ R = 1.4 \,{ \rm GeV}^{-1}\,$ 
as in the zero temperature case \cite{Boschi-Filho:2005mw} we find a critical temperature $ T_C \,\sim \,{\rm  230 MeV}\,$.

\begin{figure}
\centering
\vskip .5cm
\includegraphics[width=8cm]{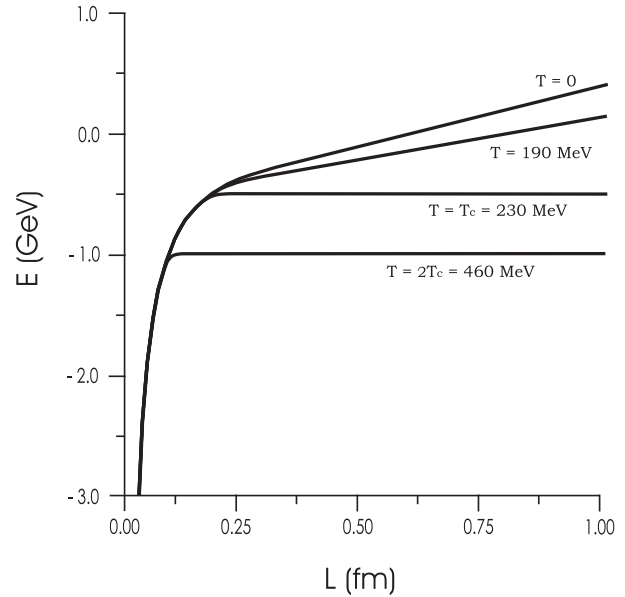}
{\caption{ Energy as a function of string end-point separation for different temperatures. } }
\end{figure}

Our results agree with lattice calculations for QCD \cite{Kaczmarek:1999mm}-\cite{Petreczky:2005bd} only for high temperatures. The results from fluctuations of strings in flat space at finite  temperature \cite{Pisarski:1982cn} and from lattice calculations \cite{deForcrand:1984cz} imply corrections to the string tension of order $\,-\, T^2\,$ at low temperatures, while our model predicts corrections of order $\,- \,T^4 \,\, $, as can be seen from  eq. (\ref{tensionT}). There is also a very recent result by Herzog \cite{Herzog:2006ra} which indicates that the dual space for temperatures below $T_C$ should be indeed the thermal AdS. 

If we had considered the thermal AdS metric for low temperatures, instead of the AdS Schwarz\-schild black hole metric, we would get no thermal corrections to the string tension. It would be interesting to find
a holographic phenomenological model that gives the expected low temperature corrections.  Let us mention the recent articles   \cite{Ghoroku:2005kg} and  \cite{Andreev:2006eh} that also discuss thermal effects in the gauge/string duality context.   
For a very recent review of thermal properties of QCD related to the quark-gluon plasma and RHIC experiment see \cite{Shuryak:2006se}. 
For other interesting results concerning gauge/string duality and QCD see for instance   \cite{Janik:1999zk}-\cite{Brodsky:2006uq}.

\bigskip
\bigskip

\noindent{\bf Acknowledgements:} We would thank Andrey Bytsenko, Sebasti\~ao A. Dias,   Jos\'e A. Helay\"el-Neto and Maria Emilia X. Guimar\~aes, organizers of ``Fifth International Conference on Mathematical Methods in Physics,'' at CBPF, Rio de Janeiro, Brazil,  where this talk was presented, for the very warm atmosphere during the workshop. The authors are partially supported by CNPq and Faperj.




\begin{thebibliography}{99}


\bibitem{Malda} J. Maldacena, Adv. Theor. Math. Phys. 2 (1998) 231.

\bibitem{GKP} S. S. Gubser , I.R. Klebanov and A.M. Polyakov, 
Phys. Lett. B428 (1998) 105.

\bibitem{Wi} E. Witten, Adv. Theor. Math. Phys. 2 (1998) 253.

\bibitem{Pe} J. L. Petersen, Int. J. Mod. Phys. A14 (1999) 3597.

\bibitem{Malda2} O. Aharony, S.S. Gubser, J. Maldacena, 
H. Ooguri and Y. Oz, Phys. Rept. 323 (2000) 183.

\bibitem{Kle} I. R. Klebanov, "TASI Lectures: Introduction to the AdS/CFT   
correspondence", hep-th 0009139.

\bibitem{FD} E. D'Hoker, D. Z. Freedman, "TASI Lectures 2001: Strings, Branes and 
Extra Dimensions",  hep-th/0201253. 

\bibitem{HOL1} G. 't Hooft, ``Dimensional reduction in quantum gravity"
in Salam Festschrifft, eds. A. Aly, J. Ellis and S. Randjbar-Daemi,
 World Scientific, Singapore, 1993, gr-qc/9310026.

\bibitem{HOL2} L. Susskind, J. Math. Phys. 36 (1995) 6377.

\bibitem{HOL3} L. Susskind and E. Witten, ``The holographic bound in anti-de 
Sitter space", SU-ITP-98-39, IASSNS-HEP-98-44, hep-th 9805114.

\bibitem{HOL4} For a covariant generalisation of the Holographic principle see:
R. Bousso, J. High Energy Phys. 06 (1999) 028;  Class. Quant. Grav. 17 (2000) 997; 
Rev. Mod. Phys. 74 (2002) 825.

\bibitem{BS} D. Bigatti and L. Susskind,  "TASI Lectures on the Holographic principle",
hep-th 0002044.

\bibitem{V} G. Veneziano, Nuovo Cim. 57A (1968) 190; Phys. Rep. C9 (1974) 199.

\bibitem{VR} M. A. Virasoro, Phys. Rev. 177 (1969) 2309; Phys. Rev. Lett. 22 (1969) 37;
Phys. Rev. D1 (1970) 2933.

\bibitem{SH} J. A. Shapiro, Phys. Rev. 179 (1969) 1345; Phys. Lett. B33 (1970) 361.

\bibitem{Planar} G. 't Hooft, Nucl. Phys. B72 (1974) 461.

\bibitem{Pol} A. M. Polyakov, Nucl. Phys. Proc. Suppl. 68 (1998) 1.

\bibitem{GSW} M. B. Green, J. H. Schwarz , E. Witten ,  "Superstring Theory", Vol. 1,
Cambridge, 1987.

\bibitem{QCD1} V. A. Matveev, R.M. Muradian and A. N. Tavkhelidze,
 Lett. Nuovo Cim. 7 (1973) 719.

\bibitem{BRO} S. J. Brodsky and G. R. Farrar, Phys. Rev. Lett 31 (1973) 1153;
Phys. Rev. D11 (1975) 1309.

\bibitem{PS} J. Polchinski and M. J. Strassler, Phys. Rev. Lett. 88 (2002) 031601.

\bibitem{BB3} H. Boschi-Filho and N. R. F. Braga, Phys. Lett. B560 (2003) 232. 

\bibitem {GI} S. B. Giddings,  Phys. Rev. D67 (2003) 126001.

\bibitem{BT} R. C. Brower , C.-I. Tan, Nucl. Phys. B662 (2003) 393.

\bibitem{AN} O. Andreev,  Phys. Rev. D67 (2003) 046001.

\bibitem{Wi2} E. Witten, Adv.Theor.Math.Phys. {\bf 2} (1998) 505.

\bibitem{Hawking:1982dh}
  S.~W.~Hawking and D.~N.~Page,
  Commun.\ Math.\ Phys.\  {\bf 87} (1983) 577.

\bibitem{MASSG} C. Csaki, H. Ooguri, Y. Oz and J. Terning, JHEP {\bf 9901} (1999) 017.

\bibitem{MASSG2} R. de Mello Koch, A. Jevicki, M. Mihailescu , J. P. Nunes,  
Phys.Rev. {\bf D58} (1998) 105009.

\bibitem{MASSG3} A. Hashimoto , Y. Oz , Nucl.Phys. {\bf B548} (1999) 167. 

\bibitem{MASSG4} C. Csaki , Y. Oz , J. Russo , J. Terning , 
Phys.Rev. {\bf D59} (1999) 065012. 

\bibitem{MASSG5} J. A. Minahan,  JHEP {\bf 9901} (1999) 020.

\bibitem{MASSG6} C. Csaki, J. Terning,  AIP Conf. Proc. {\bf 494} (1999) 321.
  
\bibitem{MASSG7} R. C. Brower, S. D. Mathur , C. I. Tan , Nucl. Phys. B587 (2000) 249. 

\bibitem{Maldacena:2000yy}
J.~M.~Maldacena and C.~Nunez,
Phys.\ Rev.\ Lett.\  {\bf 86} (2001)  588.

\bibitem{Klebanov:2000nc}
I.~R.~Klebanov and A.~A.~Tseytlin,
Nucl.\ Phys.\ B {\bf 578} (2000) 123.

\bibitem{Klebanov:2000hb}
I.~R.~Klebanov and M.~J.~Strassler,
JHEP {\bf 0008} (2000)  052.

\bibitem{Polchinski:2000uf}
J.~Polchinski and M.~J.~Strassler,
``The string dual of a confining four-dimensional gauge theory,''
arXiv:hep-th/0003136.

\bibitem{Boschi-Filho:2002vd}
  H.~Boschi-Filho and N.~R.~F.~Braga,
  JHEP {\bf 0305} (2003) 009.

\bibitem{Boschi-Filho:2002ta}
  H.~Boschi-Filho and N.~R.~F.~Braga,
  Eur.\ Phys.\ J.\ C {\bf 32} (2004) 529. 

\bibitem{deTeramond:2005su}   G.~F.~de Teramond and S.~J.~Brodsky,
Phys. Rev. Lett. {\bf 94} (2005) 201601.

\bibitem{Boschi-Filho:2005yh}
  H.~Boschi-Filho, N.~R.~F.~Braga and H.~L.~Carrion,
  Phys.\ Rev.\ D {\bf 73} (2006) 047901.

\bibitem{RY} S.~J.~Rey and J.~T.~Yee,
Eur.\ Phys.\ J.\ C {\bf 22} (2001) 379.
  
\bibitem{MaldaPRL} J.~Maldacena, 
Phys.\ Rev.\ Lett.\  {\bf 80} (1998)  4859.

\bibitem{Kinar:1998vq}
Y.~Kinar, E.~Schreiber and J.~Sonnenschein,
Nucl.\ Phys.\ B {\bf 566} (2000) 103.


\bibitem{Boschi-Filho:2005mw}
  H.~Boschi-Filho, N.~R.~F.~Braga and C.~N.~Ferreira,
  Phys.\ Rev.\ D {\bf 73} (2006) 106006.

\bibitem{Quigg:1977dd}
  C.~Quigg and J.~L.~Rosner,
  Phys.\ Lett.\ B {\bf 71} (1977) 153.

\bibitem{Eichten:1978tg}
  E.~Eichten, K.~Gottfried, T.~Kinoshita, K.~D.~Lane and T.~M.~Yan,
Phys.\ Rev.\ D {\bf 17} (1978) 3090.
[Erratum-ibid.\ D {\bf 21} (1980) 313].

\bibitem{Martin:1980jx}
  A.~Martin,
 Phys.\ Lett.\ B {\bf 93} (1980) 338.


\bibitem{Boschi-Filho:2006pe}
  H.~Boschi-Filho, N.~R.~F.~Braga and C.~N.~Ferreira,
 Phys.\ Rev.\ D {\bf 74} (2006)  086001.

\bibitem{Polchinski:1998rq} J.~Polchinski, ``String theory. Vol. {\bf 1}: An introduction to the bosonic string,'' Cambridge 1998.

\bibitem{BB2} H. Boschi-Filho and N. R. F. Braga, Phys. Lett. {\bf B525} (2002) 164.

\bibitem{PS2} J. Polchinski and M. J. Strassler, J. High Energy Phys. 05 (2003) 012. 

\bibitem{Brodsky:2003px}
  S.~J.~Brodsky and G.~F.~de Teramond,
  Phys.\ Lett.  {\bf B 582} (2004)  211.

\bibitem{AN2}
  O.~Andreev,
  Phys.\ Rev.\ {\bf D 70} (2004) 027901.

\bibitem{AN3}
  O.~Andreev,
  Phys.\ Rev.\ D {\bf 71} (2005)  066006.

\bibitem{LAT1} 
C. J. Morningstar and M. Peardon, Phys. Rev. D 56 (1997) 4043.

\bibitem{LAT2} M.J. Teper, "Physics from lattice: Glueballs in QCD; topology; SU(N) for all N ", arXiv:hep-lat/9711011.

\bibitem{Caceres:2000qe}
  E.~Caceres and R.~Hernandez,
  Phys.\ Lett.\ B {\bf 504} (2001)  64.

\bibitem{ACEP}  R. Apreda, D. E. Crooks, N. Evans, M. Petrini, 
JHEP {\bf 0405} (2004) 065.

\bibitem{Amador:2004pz}
  X.~Amador and E.~Caceres,
  JHEP {\bf 0411} (2004) 022.

\bibitem{Caceres:2005yx}
  E.~Caceres and C.~Nunez,
JHEP {\bf 0509} (2005)  027. 

\bibitem{Landshoff:2001pp}
  P.~V.~Landshoff,
  ``Pomerons,'', published in  ``Elastic and Difractive Scattering" 
 Proceedings, Ed. V. Kundrat and P. Zavada, 2002, arXiv:hep-ph/0108156.

\bibitem{Meyer:2004jc}
  H.~B.~Meyer and M.~J.~Teper,
  Phys.\ Lett.\ B {\bf 605} (2005) 344.

\bibitem{Randall:1999ee}
  L.~Randall and R.~Sundrum,
  Phys.\ Rev.\ Lett.\  {\bf 83} (1999) 3370.

\bibitem{Boschi-Filho:2004ci}
  H.~Boschi-Filho and N.~R.~F.~Braga,
  JHEP {\bf 0503} (2005)  051.

\bibitem{Policastro:2001yc}
  G.~Policastro, D.~T.~Son and A.~O.~Starinets,
  Phys.\ Rev.\ Lett.\  {\bf 87} (2001) 081601.

\bibitem{Kovtun:2004de}
  P.~Kovtun, D.~T.~Son and A.~O.~Starinets,
  Phys.\ Rev.\ Lett.\  {\bf 94} (2005) 111601.


\bibitem{Rey:1998bq}
  S.~J.~Rey, S.~Theisen and J.~T.~Yee,
  Nucl.\ Phys.\ B {\bf 527} (1998) 171.

\bibitem{Brandhuber:1998bs}
  A.~Brandhuber, N.~Itzhaki, J.~Sonnenschein and S.~Yankielowicz,
  Phys.\ Lett.\ B {\bf 434} (1998) 36.


\bibitem{Kaczmarek:1999mm}
O.~Kaczmarek, F.~Karsch, E.~Laermann and M.~Lutgemeier,
Phys.\ Rev.\ D {\bf 62} (2000) 034021.
 

\bibitem{Kaczmarek:2002mc}
O.~Kaczmarek, F.~Karsch, P.~Petreczky and F.~Zantow,
Phys.\ Lett.\ B {\bf 543} (2002) 41.
 
\bibitem{Kaczmarek:2004gv}
O.~Kaczmarek, F.~Karsch, F.~Zantow and P.~Petreczky,
Phys.\ Rev.\ D {\bf 70} (2004) 074505 
[Erratum-ibid.\ D {\bf 72} (2005) 059903].


\bibitem{Petreczky:2005bd}
  P.~Petreczky,
  Eur.\ Phys.\ J.\ C {\bf 43} (2005) 51.


\bibitem{Pisarski:1982cn}
  R.~D.~Pisarski and O.~Alvarez,
  Phys.\ Rev.\ D {\bf 26} (1982) 3735.

\bibitem{deForcrand:1984cz}
  P.~de Forcrand, G.~Schierholz, H.~Schneider and M.~Teper,
  Phys.\ Lett.\ B {\bf 160} (1985) 137.

\bibitem{Herzog:2006ra}
  C.~P.~Herzog,
  ``A holographic prediction of the deconfinement temperature,''
  arXiv:hep-th/0608151.

\bibitem{Ghoroku:2005kg}
  K.~Ghoroku and M.~Yahiro,
  Phys.\ Rev.\ D {\bf 73} (2006) 125010.

\bibitem{Andreev:2006eh}
  O.~Andreev and V.~I.~Zakharov,
   ``The Spatial String Tension, Thermal Phase Transition, and AdS/QCD,'' 
  arXiv:hep-ph/0607026.

\bibitem{Shuryak:2006se}
  E.~V.~Shuryak,
  ``Strongly coupled quark-gluon plasma: The status report,''
  arXiv:hep-ph/0608177.

\bibitem{Janik:1999zk}
  R.~A.~Janik and R.~Peschanski,
  Nucl.\ Phys.\ B {\bf 565} (2000) 193.


\bibitem{PandoZayas:2003yb}
  L.~A.~Pando Zayas, J.~Sonnenschein and D.~Vaman,
  Nucl.\ Phys.\ B {\bf 682} (2004) 3.

\bibitem{Andreev:2004sy}
  O.~Andreev and W.~Siegel,
  Phys.\ Rev.\ D {\bf 71} (2005)  086001.

\bibitem{Bigazzi:2004ze}
  F.~Bigazzi, A.~L.~Cotrone, L.~Martucci and L.~A.~Pando Zayas,
  Phys.\ Rev.\ D {\bf 71} (2005) 066002.

\bibitem{Erlich:2005qh}
  J.~Erlich, E.~Katz, D.~T.~Son and M.~A.~Stephanov,
Phys. Rev. Lett. {\bf 95} (2005) 261602.

\bibitem{DaRold:2005zs}
  L.~Da Rold and A.~Pomarol,
 Nucl.\ Phys.\  {\bf B 721} (2005) 79.


\bibitem{Evans:2005ip}
  N.~Evans, J.~P.~Shock and T.~Waterson,
Phys.Lett. {\bf B622} (2005) 165.

\bibitem{Brodsky:2005kc}
  S.~J.~Brodsky and G.~F.~de Teramond,
  AIP Conf.\ Proc.\  {\bf 814} (2006) 108.
  [arXiv:hep-ph/0510240].

\bibitem{Brodsky:2006uq}
  S.~J.~Brodsky and G.~F.~de Teramond,
  Phys.\ Rev.\ Lett.\  {\bf 96} (2006)  201601.
  







\end{thebibliography}
\end{document}